\documentclass[%
 aip,
 pof,
 amsmath,amssymb,
 reprint,%
]{revtex4-1}

\usepackage{tabularx}
\newcolumntype{C}[1]{>{\centering\arraybackslash}p{#1}}
\usepackage{dcolumn}
\usepackage{siunitx}
\usepackage{color}
\usepackage[capitalise]{cleveref}
\usepackage{tikz}
\usepackage{pgfplots}
\usepgfplotslibrary{groupplots}
\usetikzlibrary{spy,decorations.fractals,shapes.misc}
\usetikzlibrary{decorations.markings}
\usepgfplotslibrary{colormaps}
\usepgfplotslibrary{colorbrewer}
\pgfplotsset{cycle list/Dark2}
\usepackage{nicefrac}
\usepackage[caption=false]{subfig}
\usepackage{booktabs,multirow}
\usepackage{xcolor}

\makeatletter
\def\@email#1#2{%
 \endgroup
 \patchcmd{\titleblock@produce}
  {\frontmatter@RRAPformat}
  {\frontmatter@RRAPformat{\produce@RRAP{*#1\href{mailto:#2}{#2}}}\frontmatter@RRAPformat}
  {}{}
}%
\makeatother
\begin{document}

\preprint{AIP/123-QED}

\title[Shakhov-BGK for polyatomic mixtures]{A Shakhov-based Bhatnagar-Gross-Krook model for polyatomic molecules and for atomic as well as polyatomic mixtures}
\author{M. Pfeiffer}
\thanks{Corresponding author. mpfeiffer@irs.uni-stuttgart.de}
\author{F. Tuttas}
\thanks{tuttasf@irs.uni-stuttgart.de}
 \affiliation{%
   Institute of Space Systems, University of Stuttgart, Pfaffenwaldring 29, 70569
   Stuttgart, Germany
 }

\date{\today}

\begin{abstract}
The implementation of the Shakhov Bhatnagar-Gross-Krook (SBGK) method in the open-source particle code PICLas is extended for modeling of polyatomic molecules, as well as mixtures including atoms and molecules, while accounting for non-equilibrium in the internal degrees of freedom.
The conservation properties of the model are shown and the model parameter for the Prandtl number is derived.
In order to determine the viscosity and thermal conductivity of gas mixtures, the first approximation of the transport properties using collision integrals is employed.
The model is verified with simulation test cases of a supersonic Couette flow and a hypersonic flow around a 70$^\circ$ blunted cone with different flow parameters and gas compositions.
The results are compared to the Direct Simulation Monte Carlo (DSMC) method as well as the Ellipsoidal Statistical BGK (ESBGK) method to assess the accuracy of the model, where overall good agreement is achieved.
In particular, the proposed SBGK model captures the shock in front of the 70$^\circ$ blunted cone more precisely than the ESBGK model.
\end{abstract}

\maketitle

\section{Introduction}
Numerical simulations of fluid dynamics in space-related applications, as well as in microscale and nanoscale flows and vacuum technologies, pose considerable challenges, since these applications involve large density gradients, ranging from the continuum to the free molecular flow regime, which requires versatile numerical approaches.
Using the well-established Direct Simulation Monte Carlo (DSMC) method~\cite{bird}, a highly accurate solution of these multi-scale and non-equilibrium flows can be achieved.
However, the computational effort increases significantly in the transition and continuum regimes, making a coupling of DSMC with a computationally more efficient method highly desirable.

Although computational fluid dynamics (CFD) is typically used for simulations in the continuum regime, the coupling with the DSMC method is challenging due to the very different underlying approaches of both methods.
In particular, the statistical noise of the DSMC method poses a problem at the boundaries between the computational regimes of DSMC and CFD~\cite{zhang2019particle,burt-boyd-ld,pfeiffer2019evaluation,SCHWARTZENTRUBER2006402}.

In recent years, particle-based continuum methods have emerged as promising alternative, including, but not limited to, the Fokker-Planck~\cite{hossein,fp,mathiaud2016fokker,Jun2019,kim2025stochastic,kim2025particle,cui2025multiscale} and Bhatnagar-Gross-Krook (BGK)~\cite{bgk,gallis-torczynski-2,zhang2019particle,burt2006evaluation} models.
The latter approximates the Boltzmann collision integral by a relaxation process instead of modeling binary collisions as in the DSMC method.
Since with this, the mean free path and the collision frequency do not need to be resolved, the method has less restrictions regarding the choice of time step and particle weighting factor for particle simulations.
Instead, the necessary local resolution is defined by the gradients of the macroscopic moments of the distribution function present in the flow, such as density, velocity, or temperature, similar to that described in~\citet{fp}, and the time step is rather determined by the stiffness of the BGK collision term, i.e. the relaxation frequency to be resolved.
Also, different time integration and space interpolation methods can be used to achieve significantly coarser resolutions, since in comparison to DSMC, the BGK equation no longer represents a jump process.
Thus, significantly lower computational costs are expected compared to DSMC simulations, in particular for low-$Kn$ regimes.

While BGK models have been a research focus in the context of discrete velocity methods~\cite{guo2013discrete,mieussens2000discrete} for a longer time, progress has also been made with respect to particle methods~\cite{PhysRevE.106.025303, fei2021efficient, FEI2020108972, liu2020unified} in recent times.
In particular, the particle-based ellipsoidal statistical BGK (ESBGK)~\cite{esbgk} and the Shakhov BGK~\cite{shakhov} models were investigated~\cite{piclas-bgk,fei2020benchmark}, both of which are producing the correct Prandtl number of the gas compared to the Standard BGK model.

In recent years, the BGK models has been extended to diatomic and polyatomic gases~\cite{brull2025esbgkpoly,ANDRIES2000813,ANDRIES20023369,kosuge2019kinetic,rykov1975model,WANG2017237} as well as gas mixtures.
\citet{overview-bgk-mix} presents an overview of different BGK models for gas mixtures and roughly classifies them into two categories.
The first type consists of models that include a sum of multiple relaxation terms, assigning a separate relaxation term to each combination of species~\cite{asinari2008asymptotic,garzo1989kinetic,hamel1965kinetic,klingenberg2018consistent,klingenberg2018kinetic,bobylev2018general}.
This offers the advantage that inter-species collision rates, as well as the associated exchange rates of momentum and energy, can in principle be represented consistently with the Boltzmann equation.
However, the disadvantage is that the models are significantly more complex, which comes at the cost of significantly higher computational effort.
Moreover, determining free parameters, such as the different relaxation frequencies, poses additional challenges, particularly when targeting specific macroscopic properties, such as the viscosities or Prandtl numbers of the entire mixture.
These difficulties become even more pronounced for mixtures with more than two components and for polyatomic gases where internal energy exchange must also be accounted for.
The second type comprises models with a single relaxation term on the right-hand side~\cite{andries2002consistent,brull,brull2021ellipsoidal,todorova-shakhov-esbgk,todorova2020}.
A clear limitation of these models is that the correct energy and momentum exchange times cannot be captured accurately.
Nevertheless, these models offer practical advantages: they are computationally more efficient and less complex.
Macroscopic mixture properties, such as viscosity, heat flow, Prandtl number, etc., can be modeled correctly with significantly fewer parameters, whereby established mixing rules can be used.
Furthermore, since the advection part of the equation remains unchanged, the error introduced by one simplified relaxation term can be small, depending on the specific application.
For stochastic particle methods, single-relaxation-term models provide an additional advantage since only the moments of the overall mixture need to be evaluated, rather than the moments of each individual species.
As a consequence, the distribution function can be represented with a significantly smaller number of simulation particles, leading to a notable reduction in computational cost compared to multi-relaxation-term models, where the moments of all species must be resolved separately and, therefore, a larger particle count is required.

The ESBGK model has already been shown to be more efficient in the particle context~\cite{piclas-bgk, pfeiffer2019evaluation}.
However, previous comparisons indicate that the SBGK model catches the shock region in front of a body more precisely~\cite{piclas-bgk}.
Therefore, the focus of this paper is the extension of the particle-based SBGK model to polyatomic molecules as well as gas mixtures, including quantized vibrational states by using a model with only one relaxation term.
The work is based on the models of \citet{MATHIAUD202265,energy-cons} and \citet{brull,brull2021ellipsoidal}.
For ESBGK, the modeling and implementation of single-species gases, atomic, and polyatomic gas mixtures has already been presented in~\cite{bgk-poly,bgk-multispecies,bgk-poly-mix-hild} and is already widely used~\cite{FEI2026110029,10.1116/6.0004807,HE2026110219,10.1063/5.0319910}.

\section{Theory}
The BGK operator approximates the Boltzmann collision integral by modeling the relaxation of the particle distribution function
$f_s\!\left(\mathbf x,\mathbf v,t\right)$ of species $s$ at position $\mathbf x$ and velocity $\mathbf v$ toward a target distribution $f_s^t$~\cite{bgk}
\begin{equation}
	\partial_t f_s + \mathbf v \cdot \partial_{\mathbf x} f_s = \nu \left(f_s^t - f_s\right), \label{eq:BGK}
\end{equation}
where $\nu$ denotes the relaxation frequency.

In a mixed gas of atoms and molecules, the target distribution function differs for each species since molecular internal degrees of freedom (DOF) must be considered.
Atoms possess only translational DOFs, whereas molecules additionally exhibit rotational and vibrational ones.
Consequently, the target distribution for atoms contains only a translational component
\begin{equation}
	f_{s,\mathrm{Atom}}^t = f_s^{t,\mathrm{tr}}(\mathbf v),
\end{equation}
depending solely on the particle velocity $\mathbf v$.
In contrast, the molecular target distribution is factored in translational, rotational, and vibrational contributions~\cite{MATHIAUD202265,dauvois2021bgk}
\begin{equation}
	f_{s,\mathrm{Molecule}}^t = f_s^{t,\mathrm{tr}}(\mathbf v)\,
	f_s^{t,\mathrm{rot}}(E_\mathrm{rot})\,
	f_s^{t,\mathrm{vib}}(i_\mathrm{vib}),
\end{equation}
which also depend on the rotational energy $E_\mathrm{rot}$ and the vibrational quantum number $i_\mathrm{vib}$.

\subsection{Macroscopic flow quantities}
The macroscopic flow quantities for each species $s$, e.g. particle density $n$, flow velocity $\mathbf u$, and translational temperature $T_{\mathrm{tr}}$, are defined as moments of the particle distribution function:
\begin{alignat}{1}
	&n_s = \int_{\mathbb{R}^3} f_s^{\mathrm{tr}}\,\mathrm d\mathbf v,
	\quad n_s \mathbf u_s = \int_{\mathbb{R}^3} \mathbf v f_s^{\mathrm{tr}}\,\mathrm d\mathbf v, \\
	&\mathcal{E}_s = \frac{3}{2}k_\mathrm{B} T_{\mathrm{tr},s}
	= \frac{m_s}{2n_s} \int_{\mathbb{R}^3} \left| \mathbf c_s\right|^2 f_s^{\mathrm{tr}}\,\mathrm d\mathbf v,\\
	&E_s = \frac{1}{2} m_s n_s \mathbf u_s^2 + n_s \mathcal{E}_s,
\end{alignat}
where the thermal velocity is given by $\mathbf c_s = \mathbf v - \mathbf u_s$.
Here, $\mathcal{E}_s$ and $E_s$ denote the thermal and total energy of species $s$, respectively.
The macroscopic mean values of the entire flow field are then obtained as
\begin{alignat}{1}
	&n = \sum_{s=1}^M n_s, \quad m=\sum_{s=1}^M \frac{n_s}{n}m_s,\\
    &\rho = \sum_{s=1}^M m_s n_s, \quad \rho \mathbf u = \sum_{s=1}^M m_s n_s \mathbf u_s,\\
	&E = n\mathcal{E} + \frac{\rho}{2} \left|\mathbf u\right|^2 = \sum_{s=1}^M E_s,
	\quad \mathcal{E} = \frac{3}{2}k_\mathrm{B} T_{\mathrm{tr}},\\
    &p=n k_\mathrm{B} T_{\mathrm{tr}},
\end{alignat}
with the pressure of the mixture $p$.

For molecular species, the mean rotational and vibrational energies, $\left<E\right>_{\mathrm{rot},s}$ and $\left<E\right>_{\mathrm{vib},s}$, are linked to the respective rotational and vibrational temperatures $T_{\mathrm{rot},s}$ and $T_{\mathrm{vib},s}$:
\begin{alignat}{2}
	&\left<E\right>_{\mathrm{rot},s} &&= \frac{\xi_{\mathrm{rot},s}}{2}k_\mathrm{B} T_{\mathrm{rot},s}
	= \int E_\mathrm{rot} f_s^{\mathrm{rot}}\,\mathrm dE_\mathrm{rot},\\
	&\left<E\right>_{\mathrm{vib},s} &&= \sum_{j=1}^{\gamma_s} \frac{\xi_{\mathrm{vib},j,s}}{2} k_\mathrm{B} T_{\mathrm{vib},s} \nonumber
	\\
    & &&= \sum_{j=1}^{\gamma_s} \sum_{i_{\mathrm{vib},j}}
	 i_{\mathrm{vib},j} k_\mathrm{B} \Theta_{\mathrm{vib},j,s}  f_s^{\mathrm{vib}}.
\end{alignat}
Here, $\xi$ denotes the rotational and vibrational degrees of freedom.
For diatomic molecules, $\xi_\mathrm{rot}=2$; for polyatomic molecules, $\xi_\mathrm{rot}=2$ in the linear case and $\xi_\mathrm{rot}=3$ in the non-linear case~\cite{relax,herzberg1946infrared}.
The vibrational motion is modeled as a simple harmonic oscillator (SHO) with $i_{\mathrm{vib},j}$ being the quantum number of the $j$th vibrational mode, using the characteristic vibrational temperature $\Theta_\mathrm{vib}$ and defining the vibrational energy (without considering the vibrational zero-point energy) as
\begin{equation}
E_{\mathrm{vib},j,s} = i_{\mathrm{vib},j} k_\mathrm{B} \Theta_{\mathrm{vib},j,s},
\end{equation}
Due to quantization, the corresponding degrees of freedom depend on the temperature.
For a molecule with $a$ atoms, the number of vibrational modes $\gamma$ is given by~\cite{relax}:
\begin{equation}
	\gamma =
	\begin{cases}
		3a - 5, & \text{linear molecule,}\\
		3a - 6, & \text{non-linear molecule.}
	\end{cases}
\end{equation}
In the diatomic case, $\gamma = 1$.
Under the SHO assumption, the mean vibrational energy and its effective degrees of freedom can be expressed as functions of the vibrational temperature $T_{\mathrm{vib},s}$:
\begin{alignat}{1}
	&\left<E\right>_{\mathrm{vib},s}
	= \sum_{j=1}^\gamma
	\frac{k_\mathrm{B}\Theta_{\mathrm{vib},j,s}}
	{\exp(\Theta_{\mathrm{vib},j,s}/T_{\mathrm{vib},s}) - 1},\\
	&\xi_{\mathrm{vib},s}
	= \sum_{j=1}^\gamma \xi_{\mathrm{vib},j,s}
	= \sum_{j=1}^\gamma
	\frac{2\Theta_{\mathrm{vib},j,s}/T_{\mathrm{vib},s}}
	{\exp(\Theta_{\mathrm{vib},j,s}/T_{\mathrm{vib},s}) - 1}.
\end{alignat}

\subsection{Shakhov BGK model for polyatomic molecules and mixtures}
The Shakhov BGK model produces the correct Prandtl number by an adaptation of the heat flux in the target distribution
\begin{alignat}{1}
    f^\text{SBGK,tr}_s &= f^\text{M}_s \left[ 1 + \left(1-\alpha Pr\right) \frac{\mathbf{c^\mathrm{T}q}}{5n\left(k_\text{B}T_{\text{tr,rel}}\right)^2/m_s}\right.\nonumber\\
    &\left. \times\left(\frac{\left|\mathbf{c}\right|^2}{ k_\text{B}T_{\text{tr,rel}}/m_s} - 5\right) \right] \\    \label{eq:sbgk}
    f^\text{M}_s &= \frac{n_s}{(2\pi k_\mathrm{B} T_{\text{tr,rel}}/m_s)^{3/2}}\exp{\left(-\frac{\left|\mathbf{c}\right|^2}{2k_\mathrm{B} T_{\text{tr,rel}}/m_s}\right)}
\end{alignat}
with the heat flux vector
\begin{alignat}{1}
	\mathbf q &=\sum_{s=1}^M \int \mathbf c \left[ \frac{m_s}{2}\left|\mathbf c \right|^2f_s^{\mathrm{tr}} +\int E_\mathrm{rot}f_s^{\mathrm{tr}}f_s^{\mathrm{rot}}\,\text d E_\mathrm{rot} \right. \nonumber\\
    &\left.+ \sum_{j=1}^\gamma \sum_{i_{\mathrm{vib},j}}
	i_{\mathrm{vib},j} k_\mathrm{B} \Theta_{\mathrm{vib},j,s}f_s^{\mathrm{tr}}f_s^{\mathrm{vib}}\right]\text d\mathbf v,
\end{alignat}
the thermal velocity of the mixture $\mathbf{c}=\mathbf{v}-\mathbf{u}$, and the Maxwellian distribution $f^\text{M}_s$. The parameter $Pr$ represents the target Prandtl number of the gas mixture, which can be obtained using Wilke’s mixing rule~\cite{wilke} or collision integrals~\cite{collint} (see Section~\ref{sec:mixture}).
The parameter $\alpha$ is a variable of the model depending on the mass fraction, density fraction, and internal degrees of freedom~\cite{brull2021ellipsoidal}:
\begin{equation}
    \alpha = m \frac{\sum_{s=1}^M \frac{n_s}{m_s} \left(5 + \xi_{\mathrm{int},s}\right)}{\sum_{s=1}^M n_s \left(5 + \xi_{\mathrm{int},s}\right)}.\label{eq:alpha}
\end{equation}
The internal degrees of freedom are calculated as the sum of the vibrational and rotational degrees of freedom, respectively:
\begin{equation}
	\xi_{\mathrm{int},s} = \xi_{\mathrm{vib},s}+\xi_{\mathrm{rot},s}.
	\label{eq:intDOF}
\end{equation}
The model presented in this work employs a single relaxation term for each species~$s$.
It satisfies the indifferentiability principle, ensuring that the formulation reduces to the single-species case when all species are identical.
In thermodynamic equilibrium, the model reproduces the Maxwellian distribution.
The relaxation frequency~$\nu$ is defined as
\begin{equation}
	\nu = \frac{n k_{\mathrm{B}} T_{\mathrm{tr}}}{\mu}\, \label{eq:nu}
\end{equation}
where $\mu$ denotes the mixture viscosity and $T_{\mathrm{tr}}$ the translational temperature.

For molecular species, the rotational and vibrational components of the distribution function must also be specified.
They are adopted directly from \citet{MATHIAUD202265} and read as:
\begin{alignat}{1}
	f_s^{t,\mathrm{rot}} &=
	\frac{E_\mathrm{rot}^{(\xi_{\mathrm{rot},s}-2)/2}}
	{\Gamma(\xi_{\mathrm{rot},s}/2)\,\left(k_\mathrm{B} T_{\mathrm{rot,rel},s}\right)^{\xi_{\mathrm{rot},s}/2}}
	\nonumber\\
    &\times\exp\!\left[-\frac{E_\mathrm{rot}}{k_\mathrm{B} T_{\mathrm{rot,rel},s}}\right], \label{eq:distrot}\\
	f_s^{t,\mathrm{vib}} &= \prod_{j=1}^{\gamma_s} f_{s,j}^{\mathrm{t,vib}}(i_j)=
	\prod_{j=1}^{\gamma_s}
	\left(1-\exp\!\left[-\frac{\Theta_{\mathrm{vib},j,s}}{T_{\mathrm{vib,rel},s}}\right]\right)\nonumber\\
	&\times\exp\!\left[-i_j\frac{\Theta_{\mathrm{vib},j,s}}{T_{\mathrm{vib,rel},s}}\right], \label{eq:distvib}
\end{alignat}
where $T_{\mathrm{tr,rel}}$, $T_{\mathrm{rot,rel},s}$, and $T_{\mathrm{vib,rel},s}$ denote the effective relaxation temperatures of the translational, rotational, and vibrational modes.
These temperatures are chosen such that the energy relaxation follows the Landau-Teller form:
\begin{alignat}{1}
	\frac{\mathrm{d}\!\left<E\right>_{r,s}(T_{r,s})}{\mathrm{d}t}
	&= \frac{1}{Z_{r,s}\tau_{s,\mathrm{C}}}
	\left(\left<E\right>_{r,s}(T_{\mathrm{tr}}) - \left<E\right>_{r,s}(T_{r,s})\right),\nonumber\\
	r &= \mathrm{rot},\,\mathrm{vib}.
\end{alignat}
Here, $Z_{r,s}$ represents the collision number of the internal degree of freedom $r$ for species $s$,
and $\tau_{s,\mathrm{C}} = 1/\nu_{s,\mathrm{C}}$ denotes the characteristic collision time.
It should be noted that $\tau_{s,\mathrm{C}}$ differs from $\tau = 1/\nu$,
the relaxation time used in the BGK operator (Eq.~\eqref{eq:BGK}).
A more detailed discussion can be found in \citet{MATHIAUD202265}.

The collision frequency for the Variable Hard Sphere (VHS) model for species~$s$
in a mixture of $M$ species is given by~\cite{chapman-cowling}
\begin{equation}
	\nu_{s,\mathrm{C}}
	= \sum_{k=1}^{M}
	2 d_{s,k}^2 n_k
	\sqrt{\frac{m_s + m_k}{m_s m_k} \, 2\pi k_\mathrm{B} T_{s,k}}
	\left(\frac{T_{\mathrm{tr}}}{T_{s,k}}\right)^{1-\omega_{s,k}}.
\end{equation}
The effective VHS parameters are evaluated as
\begin{align}
	d_{s,k} &= \tfrac{1}{2}\left(d_{\mathrm{VHS},s} + d_{\mathrm{VHS},k}\right), \\
	T_{s,k} &= \tfrac{1}{2}\left(T_{\mathrm{VHS},s} + T_{\mathrm{VHS},k}\right), \\
	\omega_{s,k} &= \tfrac{1}{2}\left(\omega_{\mathrm{VHS},s} + \omega_{\mathrm{VHS},k}\right).
\end{align}
If direct VHS data are available for specific collision pairs,
they can be used instead of the averaged values~\cite{swaminathan2016recommended,10.1063/5.0118040,hong2023optimized}.
Based on these definitions, the effective relaxation temperatures
$T_{\mathrm{tr,rel}}$, $T_{\mathrm{rot,rel},s}$, and $T_{\mathrm{vib,rel},s}$ are obtained as
\begin{alignat}{1}
	&\left<E\right>_{r,s}(T_{r,\mathrm{rel},s}) =
	\left<E\right>_{r,s}(T_{r,s}) \nonumber \\
	& \quad+ \frac{\tau}{Z_{r,s}\tau_{s,\mathrm{C}}}
	\left(\left<E\right>_{r,s}(T_{\mathrm{tr}}) - \left<E\right>_{r,s}(T_{r,s})\right),
	\label{eq:newtempinner}\\
	&\left<E\right>_{\mathrm{tr}}(T_{\mathrm{tr,rel}}) =
	\left<E\right>_{\mathrm{tr}}(T_{\mathrm{tr}}) \nonumber \\
	& \quad- \sum_{r}\sum_{s=1}^{M_\mathrm{molec}}
	\frac{\tau}{Z_{r,s}\tau_{s,\mathrm{C}}}
	\left(\left<E\right>_{r,s}(T_{\mathrm{tr}}) - \left<E\right>_{r,s}(T_{r,s})\right).
    \label{eq:newtemptrans}
\end{alignat}

\subsubsection{Conservation properties of the model}
First, note the following integral properties for the target distribution functions:
\begin{alignat}{1}
&\sum_{s=1}^M m_s\int_{\mathbb{R}^3} f^\text{SBGK,tr}_s(\mathbf{v}) \, \mathrm{d}\mathbf{v}
=\sum_{s=1}^M \rho_s= \rho, \label{eq:dens}\\
&\sum_{s=1}^M m_s\int_{\mathbb{R}^3} \mathbf{v} \, f^\text{SBGK,tr}_s(\mathbf{v}) \, \mathrm{d}\mathbf{v}
=\sum_{s=1}^M  \rho_s \mathbf{u}_s = \rho \mathbf{u}, \label{eq:momentum}\\
&\sum_{s=1}^M m_s \int_{\mathbb{R}^3} \mathbf{c} \mathbf{c}^\mathrm{T}
\, f^\text{SBGK,tr}_s(\mathbf{v}) \, \mathrm{d}\mathbf{v}
= \rho k_\mathrm{B} T_{\text{tr,rel}}\mathrm I, \label{eq:presstens}\\
&\sum_{s=1}^M \int \mathbf c \left[ \frac{m_s}{2}\left|\mathbf c \right|^2f_s^{\mathrm{SBGK,tr}} \vphantom{\sum_{j=1}^\gamma}\right. \nonumber\\
&\quad + \int E_\mathrm{rot}f_s^{\mathrm{SBGK,tr}}f_s^{t,\mathrm{rot}}\,\text d E_\mathrm{rot} \nonumber\\
&\left.\quad+ \sum_{j=1}^\gamma \sum_{i_{\mathrm{vib},j}} i_{\mathrm{vib},j} k_\mathrm{B} \Theta_{\mathrm{vib},j,s}f_s^{\mathrm{SBGK,tr}}f_s^{t,\mathrm{vib}}\right]\text d\mathbf v \nonumber\\
&\quad= (1-\alpha Pr)\mathbf{q},
\label{eq:tr_moments}
\end{alignat}
\begin{alignat}{1}
&\int_{0}^{\infty} f_s^{t,\mathrm{rot}}(E_{\mathrm{rot}}) \, \mathrm{d} E_{\mathrm{rot}}
= 1, \\
&\int_{0}^{\infty} E_{\mathrm{rot}} \, f_s^{t,\mathrm{rot}}(E_{\mathrm{rot}}) \, \mathrm{d}E_{\mathrm{rot}}
= \left<E\right>_{\mathrm{rot},s}\!\left(T_{\mathrm{rot},\mathrm{rel},s}\right),
\label{eq:rot_moments}
\end{alignat}
\begin{alignat}{1}
&\sum_{i_1,\dots,i_{\gamma_s}=0}^{\infty} f_{s}^{t,\mathrm{vib}}(i_j)
= 1, \\
&\sum_{i_1,\dots,i_{\gamma_s}=0}^{\infty} \left(\sum_{j=1}^{\gamma_s}i_{j} k_\mathrm{B} \Theta_{\mathrm{vib},j,s}\right)f_{s}^{t,\mathrm{vib}}(i_j) \nonumber\\
&\qquad \qquad= \left<E\right>_{\mathrm{vib},s}\!\left(T_{\mathrm{vib},\mathrm{rel},s}\right).
\label{eq:vib_moments}
\end{alignat}
Mass conservation and momentum conservation are direct consequences of the integrals \eqref{eq:dens} and \eqref{eq:momentum}.
Energy conservation follows from a combination of Eqs. \eqref{eq:presstens}, \eqref{eq:rot_moments}, and \eqref{eq:vib_moments}, as well as the conditions of Eqs. \eqref{eq:newtempinner} and \eqref{eq:newtemptrans}, as shown in \citet{MATHIAUD202265} for one species.

\subsubsection{Prandtl number of the model}
As shown in \citet{shakhov}, the Shakhov model does not modify the second velocity moment and therefore does not affect the viscous stress tensor.  Therefore, the viscous stress tensor $\sigma$ of the mixture is given as the sum over all
species and coincides with that of the BGK model. However, as shown in~\citet{MATHIAUD202265,10.1063/1.3640083}, it
contains an additional volume viscosity contribution
$\zeta\,(\nabla\cdot \mathbf u)\, \mathbf I$ arising from internal degrees of freedom, where
$\zeta$ denotes the volume viscosity coefficient:
\begin{equation}
\sigma=\mu\left(\nabla \mathbf u + (\nabla \mathbf u)^{\mathrm T}-\frac{2}{3} (\nabla\cdot \mathbf u)\, \mathbf I\right)+\zeta\, (\nabla\cdot \mathbf u)\, \mathbf I .
\end{equation}
As a consequence, the dynamic viscosity of the mixture in the present model is
simply given by
\begin{equation}
\mu = \frac{p}{\nu}.\label{eq:mu}
\end{equation}
The heat flux is defined as
\begin{equation}
q = \sum_{s=1}^M \int B_s(\mathbf v) f_s^{(1)}\, \mathrm d\mathbf v
\end{equation}
with the definition of the Chapman--Enskog expansion,
\begin{equation}
f_s = M_s + \frac{Kn}{\nu} f_s^{(1)} + \mathcal{O}((Kn/\nu)^2),
\end{equation}
where $M_s$ denotes the Maxwellian distribution associated with species $s$ and $Kn$ is the Knudsen number.
The first-order correction $f_s^{(1)}$ is expanded in terms of the standard Sonine
polynomials corresponding to the deviatoric stress tensor and the heat flux, whereas the Sonine polynom needed for the heat flux is given as:
\begin{alignat}{1}
B_s( \mathbf  v, &E_\mathrm{rot}, i_\mathrm{vib})=(\mathbf v-\mathbf u)
\left(\frac12 m_s |\mathbf v- \mathbf u|^2-\frac52 k_\mathrm{B} T \right.\\
&+ E_{\mathrm{rot},s} - \frac{\xi_{\mathrm{rot},s}}{2}k_\mathrm{B} T \\
&\left.+ \sum_{j=1}^\gamma  i_{\mathrm{vib},j} k_\mathrm{B} \Theta_{\mathrm{vib},j,s} - \frac{\xi_{\mathrm{vib},s}}{2}k_\mathrm{B} T
\right).
\end{alignat}
In the Shakhov model, the heat flux consists of two contributions: a BGK contribution driven by the temperature gradient and a Shakhov correction proportional to the total heat flux itself.
Following the computations performed in \citet{brull2021ellipsoidal}, the Sonine integrals are explicitly known and yield
\begin{alignat}{1}
q = &- \frac{1}{\nu}\left( k_\mathrm{B}^3 T^3
\sum_{s=1}^M \left(\frac{5}{2}+\frac{\xi_{\mathrm{rot},s}}{2}+\frac{\xi_{\mathrm{vib},s}}{2}\right)\frac{n_s}{m_s}
\right)\frac{\nabla T}{k_\mathrm{B}T^2}\\
&+(1-Pr_\mathrm{S})\, q,
\end{alignat}
where $Pr_\mathrm{S}$ denotes the Shakhov model parameter.
Solving the resulting self-consistency relation leads to Fourier's law
\begin{equation}
    q = -\kappa \nabla T
\end{equation}
with thermal conductivity
\begin{equation}
\kappa=\frac{p}{\nu\,Pr_\mathrm{S}}k_\mathrm{B}\sum_{s=1}^M \left(\frac{5}{2}+\frac{\xi_{\mathrm{rot},s}}{2}+\frac{\xi_{\mathrm{vib},s}}{2}\right) \frac{n_s}{m_s}.\label{eq:kappa}
\end{equation}
The Prandtl number of the mixture is defined by
\begin{equation}
Pr_\mathrm{phys}=\frac{\mu c_\mathrm{p}}{\kappa},
\end{equation}
with the specific heat capacity
\begin{equation}
c_\mathrm{p}=\frac{k_\mathrm{B}}{\rho}\sum_{s=1}^M \left(\frac{5}{2}+\frac{\xi_{\mathrm{rot},s}}{2}+\frac{\xi_{\mathrm{vib},s}}{2}\right) n_s
\label{eq:cp}
\end{equation}
Substituting the expressions for $\mu$ and $\kappa$ as well as $\alpha$ (Eqs.~\eqref{eq:mu},\eqref{eq:kappa},\eqref{eq:alpha}) gives
\begin{equation}
Pr_{\mathrm{phys}}=\frac{Pr_\mathrm{S}}{\alpha}.
\end{equation}
Hence, to reproduce a prescribed physical Prandtl number
$Pr_{\mathrm{phys}}$ of the mixture, the Shakhov parameter must be chosen as
\begin{equation}
Pr_\mathrm{S}=\alpha Pr_{\mathrm{phys}}.
\end{equation}

\subsubsection{Single species gas properties}
In the case where no gas mixture is present and a single polyatomic gas is simulated,
the described model simplifies accordingly and $\alpha=1$ applies.
The viscosity required for the relaxation frequency $\nu$ is determined using
the well-known exponential law for the viscosity $\mu$ using the VHS model:
\begin{equation}
    \mu = \mu_{\mathrm{ref}}\left(\frac{T}{T_{\mathrm{VHS}}}\right)^{\omega_{\mathrm{VHS}}},
\end{equation}
Furthermore, for a VHS gas, the reference dynamic viscosity is calculated from the reference particle diameter $d_{\mathrm{VHS}}$ as
\begin{equation}
    \mu_{\mathrm{ref}} =
    \frac{30\sqrt{m k_{\mathrm{B}} T_{\mathrm{VHS}}}}
    {\sqrt{\pi}\,4(5 - 2\omega_{\mathrm{VHS}})(7 - 2\omega_{\mathrm{VHS}})d_{\mathrm{VHS}}^{2}}.
\end{equation}
The Prandtl number of a molecular gas depends on its internal degrees of freedom
and can be evaluated for a single-species case as
\begin{equation}
    Pr = \frac{2\left(5 + \xi_{\mathrm{rot}} + \xi_{\mathrm{vib}}\right)}
    {15 + 2\left(\xi_{\mathrm{rot}} + \xi_{\mathrm{vib}}\right)}.
\end{equation}

\subsection{Gas mixture properties} \label{sec:mixture}
When applying the Shakhov model to gas mixtures, an accurate determination of the Prandtl number is likewise required to evaluate the target distribution of Eq.~\eqref{eq:sbgk} as well as the viscosity of the gas to define the relaxation frequency of Eq.~\eqref{eq:nu}.
This becomes significantly more complex than in the case of a single species.
In the present model, any suitable source or method can be used to evaluate the viscosity and thermal conductivity of the mixture.
The same challenges appear in multi-species CFD solvers, where identical solution strategies are commonly applied.
Typically, transport properties are determined using either Wilke’s mixture rule~\cite{wilke} or the first-approximation method based on collision integrals~\cite{collint}, with several computational libraries available for this purpose~\cite{Scoggins2020,kee1986fortran,ern2004eglib}.

In this work, Wilke’s rule is not discussed in detail, as the first-approximation approach has been shown to yield more accurate results for the considered cases (see Ref.~\cite{bgk-poly-mix-hild}).
Nevertheless, a comprehensive description of Wilke’s method can be found in~\citet{bgk-poly-mix-hild}, where both approaches are compared in detail.
To ensure consistent transport coefficients with the DSMC simulations, the procedure for the VHS potential model is adopted in this paper.

In the first approximation of transport properties, the viscosity of each species~$s$ is expressed using the collision integral~$\Omega_s^{(2)}(2)$~\cite{collint}:
\begin{equation}
    \mu_s = \frac{5 k_{\mathrm{B}} T_{\mathrm{tr},s}}{8\, \Omega_s^{(2)}(2)}.
\end{equation}
The total mixture viscosity is obtained from the species contributions~$b_s$ with
\begin{equation}
    \mu = \sum_{s=1}^{M} b_s,
\end{equation}
which are determined by solving the coupled system described in~\citet{chapman-cowling}
\begin{align}
	\chi_s &= b_s\left(\frac{\chi_s}{\mu_s} + \sum_{k\neq s}\frac{3\chi_s}{(\rho_k'+\rho_s')D_{sk}}\left(\frac{2}{3}+\frac{m_k}{m_s}A_{sk}\right)\right)\nonumber\\
	&-\chi_s\sum_{k\neq s}\frac{3 b_k}{(\rho_k'+\rho_s')D_{sk}}\left(\frac{2}{3}-A_{sk}\right)
\end{align}
with the mole fraction $\chi$, and the density $\rho_k'$ of species $k$ when pure at pressure and temperature of the actual gas mixture.
The binary diffusion coefficient and auxiliary parameter are given by
\begin{align}
    D_{sk} &= \frac{3 k_{\mathrm{B}} T_{\mathrm{tr}}}{16 n m^*_{sk} \Omega_{sk}^{(1)}(1)}, \\
    A_{sk} &= \frac{\Omega_{sk}^{(2)}(2)}{5\, \Omega_{sk}^{(1)}(1)}.
\end{align}
Similarly, the mixture thermal conductivity is computed as
\begin{equation}
    \kappa = \sum_{s=1}^{M} a_s
    + \sum_{s=1}^{M_\mathrm{molec}} (\kappa_{\mathrm{rot},s} + \kappa_{\mathrm{vib},s}),
\end{equation}
where $a_s$ denotes the translational contribution of species~$s$, while
$\kappa_{\mathrm{rot},s}$ and $\kappa_{\mathrm{vib},s}$ represent the rotational and vibrational parts, respectively.
Factors $a_s$ are determined by solving the system
\begin{align}
	\chi_s &= a_s\left[\frac{\chi_s}{\kappa_s} + \sum_{k\neq s}\frac{\chi_k}{5k_{\mathrm{B}} n D_{sk}} \cdot\left(6\left(\frac{m_s}{m_k + m_s}\right)^2\right.\right.\nonumber\\
	&\left.\left. + (5-4B_{sk})\left(\frac{m_k}{m_k + m_s}\right)^2 + 8\frac{m_k m_s}{(m_s+m_k)^2}A_{sk}\right)\right]\nonumber\\
	&-\chi_s\sum_{k\neq s}a_k\frac{m_k m_s}{(m_s+m_k)^2}(5k_{\mathrm{B}}n D_{sk})^{-1}\nonumber \\
	&\cdot\left(11-4B_{sk}-8A_{sk}\right).
\end{align}
The first approximation of the thermal conductivity of a species is~\cite{chapman-cowling}
\begin{equation}
    \kappa_s = \frac{25\, c_{\mathrm{v,tr},s}\, k_{\mathrm{B}} T_{\mathrm{tr},s}}{16\, \Omega_s^{(2)}(2)},
    \quad c_{\mathrm{v,tr},s} = \frac{3 k_{\mathrm{B}}}{2 m_s}.
\end{equation}
and the parameter $B_{sk}$ is defined by
\begin{equation}
	B_{sk}=\frac{5\Omega_{sk}^{(1)}(2)-\Omega_{sk}^{(1)}(3)}{5\Omega_{sk}^{(1)}(1)}.
\end{equation}
For the VHS model, the relevant collision integrals and the parameter $B_{sk}$ are given by~\cite{Stephani2012}:
\begin{alignat}{2}
    &\Omega_{sk}^{\mathrm{VHS},(1)}(1) =\,
    &&\frac{\pi d_{\mathrm{VHS}}^2}{2}
    \sqrt{\frac{k_{\mathrm{B}} T_{\mathrm{tr}}}{2\pi m^*_{sk}}}
    \left(\frac{T_{\mathrm{VHS}}}{T_{\mathrm{tr}}}\right)^{\omega_{\mathrm{VHS}} - 1/2} \nonumber\\
    & &&\frac{\Gamma(7/2 - \omega_{\mathrm{VHS}})}{\Gamma(5/2 - \omega_{\mathrm{VHS}})}, \\
    &\Omega_{sk}^{\mathrm{VHS},(2)}(2) = \,
    &&\frac{\pi d_{\mathrm{VHS}}^2}{3}
    \sqrt{\frac{k_{\mathrm{B}} T_{\mathrm{tr}}}{2\pi m^*_{sk}}}
    \left(\frac{T_{\mathrm{VHS}}}{T_{\mathrm{tr}}}\right)^{\omega_{\mathrm{VHS}} - 1/2} \nonumber\\
	& &&\frac{\Gamma(9/2 - \omega_{\mathrm{VHS}})}{\Gamma(5/2 - \omega_{\mathrm{VHS}})},
\end{alignat}
\begin{equation}
	B^{\mathrm{VHS}}_{sk} =\frac{5\Gamma(9/2-\omega_{\mathrm{VHS}})-\Gamma(11/2-\omega_{\mathrm{VHS}})}{5\Gamma(7/2-\omega_{\mathrm{VHS}})}.
\end{equation}
The rotational and vibrational contributions to the thermal conductivity are approximated by
\begin{equation}
    \kappa_{r,s} = \frac{n_s m_s c_{\mathrm{v},r,s}}
    {\sum_{k=1}^{M} \chi_k D_{sk}^{-1}},
    \quad c_{\mathrm{v},r,s} = \frac{\xi_{r,s} k_{\mathrm{B}}}{2 m_s},
    \quad r = \mathrm{rot,vib}.
\end{equation}

\subsection{Implementation}
The implementation of the proposed method is done in the open-source code PICLas~\cite{piclas} using a particle method.
The fundamental idea behind the implementation of the particle method is the same as that presented in~\citet{bgk-poly-mix-hild} for the ESBGK implementation.
However, since the focus of the present work lies on the extension of the Shakhov method rather than on the particle method itself, details regarding the computation of internal energies, as well as energy and momentum conservation, and the general concept of the stochastic particle approach, are referred to in Ref.~\cite{bgk-poly-mix-hild}.
The key difference in the implementation of both methods is the sampling process of new translational states from the translational target distribution functions.
For the Shakhov BGK model, an acceptance--rejection sampling technique is used, as described in Ref.~\cite{piclas-bgk}.

\section{Simulation Results}
The SBGK model previously described is validated using simulation test cases of a supersonic Couette flow with different gas compositions and a hypersonic flow over a 70$^\circ$ blunted cone with varying conditions.
The species-specific VHS parameters are provided in Table~\ref{t:VHS}.
For collisions between unlike species, the parameters are determined by averaging the respective species values (collision-averaged approach).
\begin{table}
	\centering
	\renewcommand{\arraystretch}{1.3}
	\begin{tabular}{l|l|l|l}
		Gas species & $d_\mathrm{VHS}$ / $\si{\meter}$ & $T_\mathrm{VHS}$ / $\si{\kelvin}$ & $\omega_\mathrm{VHS}$ / $-$ \\
		\hline
		$\mathrm{He}$ & $\num{2.33e-10}$ & $273$ & $0.77$ \\
		$\mathrm{Ar}$ & $\num{4.05e-10}$ & $273$ & $0.77$ \\
		$\mathrm{N}$ & $\num{3.00e-10}$ & $273$ & $0.74$ \\
		$\mathrm{N}_2$ & $\num{4.17e-10}$ & $273$ & $0.74$ \\
		$\mathrm{O}_2$ & $\num{3.98e-10}$ & $273$ & $0.74$ \\
		$\mathrm{CO}_2$ & $\num{5.10e-10}$ & $273$ & $0.74$ \\
	\end{tabular}
	\caption{VHS species-specific parameters~\cite{bird}.}
	\label{t:VHS}
\end{table}

\subsection{Supersonic Couette flow}
For the supersonic Couette flow, three different cases were investigated in order to demonstrate the functionality of the model for different gas compositions.
First, a pure nitrogen case ($\mathrm{N_2}$) was considered, second, an atomic mixture consisting of argon and helium (Ar--He) was analyzed, and third, a mixed atomic-molecular case comprising nitrogen molecules and nitrogen atoms ($\mathrm{N_2}$--N) was examined.
The exact composition and the corresponding density values for each case are listed in Table~\ref{t:couette}.
\begin{table}
	\centering
	\renewcommand{\arraystretch}{1.3}
	\begin{tabular}{l|l|l|l}
		& Gas species & Composition & $n_0$ / $\si{\per\cubic\meter}$ \\
		\hline
		Case 1 & $\mathrm{N}_2$ & $100\:\%$ & $\num{1.3e20}$ \\
		Case 2 & $\mathrm{Ar}$-$\mathrm{He}$ & $50\:\%$-$50\:\%$ & $\num{1.3e20}$  \\
		Case 3 & $\mathrm{N}_2$-$\mathrm{N}$ & $50\:\%$-$50\:\%$ & $\num{1.3e20}$ \\
	\end{tabular}
	\caption{Overview of Couette flow test cases.}
\label{t:couette}
\end{table}
In all simulations, the wall temperature was set to $T_{\mathrm{wall}} = 273\,\mathrm{K}$.
The upper and lower walls moved with velocities of $v_{\mathrm{wall}} = \pm 350\,\mathrm{ms^{-1}}$, respectively.
An accurate representation of the Prandtl number is essential for correctly predicting the temperature field in the Couette flow.

\begin{figure}\centering
  \includegraphics[width=0.9\linewidth]{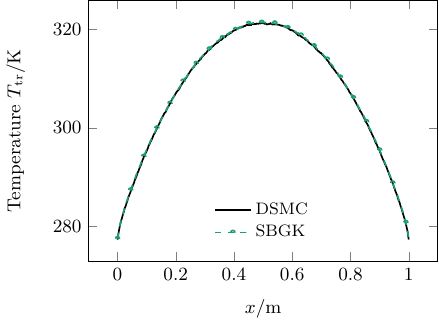}
  \caption{Stationary translational temperature profile of the N$_2$ Couette flow.}\label{fig:couetteN2}
\end{figure}

In the pure $\mathrm{N_2}$ case, it is particularly important to properly account for the internal degrees of freedom in the evaluation of the Prandtl number, ensuring the correct relaxation behavior of both the pressure tensor and the heat flux, as well as an accurate treatment of the internal energy modes themselves.
For the pure $\mathrm{N_2}$ flow, very good agreement is obtained for both translational and rotational temperature, as shown in Fig.~\ref{fig:couetteN2} for the translational temperature by way of example.

\begin{figure}\centering
  \includegraphics[width=0.9\linewidth]{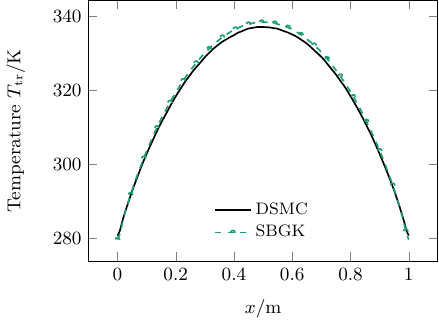}
  \caption{Stationary translational temperature profile of the Couette flow with Ar--He mixture.}\label{fig:couetteAr-He}
\end{figure}

In the second case, the Ar--He mixture, the Prandtl number depends on the accurate computation of the viscosity and thermal conductivity of the mixture, as described in Section~\ref{sec:mixture}.
Figure~\ref{fig:couetteAr-He} presents the translational temperature profile of the simulation of the Ar--He mixture.
In general, the SBGK results obtained using collision integrals (see Section~\ref{sec:mixture}) show very good agreement with the DSMC results.
However, in this case, a slight deviation is observed, which was not present in the pure $\mathrm{N_2}$ flow.
This discrepancy is likely caused by the large mass ratio between argon and helium, leading to a less accurate approximation of the effective Prandtl number. This was also already discussed in the corresponding ESBGK paper~\cite{bgk-poly-mix-hild}.

This interpretation is further supported by the third case, the $\mathrm{N_2}$--N mixture.
Although this configuration is more complex because it involves both a mixture and a molecular species with internal degrees of freedom, the agreement between SBGK and DSMC is nearly perfect.
In contrast to the Ar--He case, the mass ratio between $\mathrm{N_2}$ and N is significantly smaller, which appears to improve the precision of the Prandtl number approximation.
\begin{figure}\centering
  \includegraphics[width=0.9\linewidth]{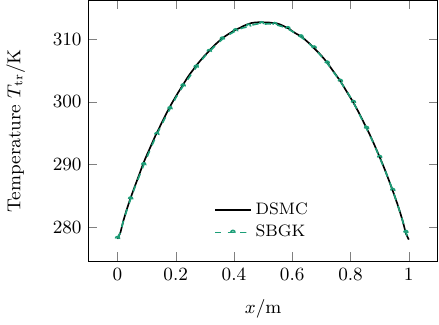}
  \caption{Stationary translational temperature profile of the Couette flow with $\mathrm{N_2}$--N mixture.}\label{fig:couetteN2-N}
\end{figure}

\subsection{70$^{\circ}$ blunted cone}
\begin{figure}
	\centering
	\includegraphics[width=0.75\linewidth]{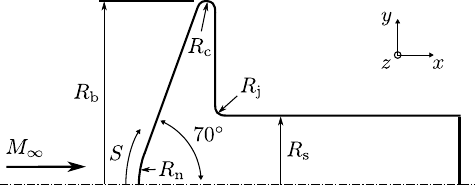}
	\caption{Geometry of the $70^\circ$ blunted cone. $R_{\mathrm{b}}=\SI{25.0}{\milli\meter}$, $R_{\mathrm{c}}=\SI{1.25}{\milli\meter}$, $R_{\mathrm{j}}=\SI{2.08}{\milli\meter}$, $R_{\mathrm{n}}=\SI{12.5}{\milli\meter}$, $R_{\mathrm{s}}=\SI{6.25}{\milli\meter}$. $S$ denotes the arc length along the surface.}
	\label{fig:70cone_geometry}
\end{figure}

To assess the performance of the SBGK mixture model, particularly in the presence of strong gradients such as those occurring across shocks, simulations of a hypersonic flow over a 70$^{\circ}$ blunted cone were conducted.
The cone geometry is depicted in Fig.~\ref{fig:70cone_geometry}.
All simulations were performed in a two-dimensional axisymmetric configuration.
At the cone surface, diffuse reflection with complete thermal accommodation was assumed, prescribing a constant wall temperature of $T_\mathrm{wall}=\SI{300}{K}$.
The corresponding inflow conditions are summarized in Table~\ref{t:70cone_inflow}.
\begin{table}
	\centering
	\small
	\renewcommand{\arraystretch}{1.3}
	\begin{tabular}{l|l|l|l}
		& Case 1 & Case 2 & Case 3 \\
		\hline
		Gas species & $\mathrm{N}_2$ & $\mathrm{CO}_2$--$\mathrm{N}_2$ & $\mathrm{N}_2$--$\mathrm{O}_2$--$\mathrm{N}$  \\
		$T_\infty$ / $\si{\kelvin}$ & $13.58$ & $13.3$ & $13.3$  \\
		$u_\infty$ / $\si{\meter\per\second}$ & $1502.4$ & $1502.57$ & $1502.57$ \\
		$n_\infty$ / $\si{\per\cubic\meter}$ & $1.11\cdot 10^{21}$ & $3.7\cdot 10^{20}$ & $5.55\cdot 10^{20}$ \\
        Composition & 1 & 0.5-0.5 & $\nicefrac{1}{3}$-$\nicefrac{1}{3}$-$\nicefrac{1}{3}$\\
	\end{tabular}
	\caption{Inflow conditions of hypersonic flow around a 70$^{\circ}$ blunted cone.}
	\label{t:70cone_inflow}
\end{table}

The first test case again considers a pure $\mathrm{N_2}$ flow in order to evaluate the behavior of the model under hypersonic conditions.
An advantage of this configuration is the availability of both DSMC reference data and experimental surface heat flux measurements for validation.
In the second case, a $\mathrm{CO_2}$--$\mathrm{N_2}$ mixture was investigated to evaluate the accuracy of the model for a polyatomic-diatomic mixture with a larger number of internal degrees of freedom, particularly vibrational modes.
Such conditions are representative, for example, of a Mars entry scenario.
Finally, a third case including three species with internal degrees of freedom was selected to examine the performance of the model for mixtures with more than two components.

\subsubsection{Case 1: N\textsubscript{2}}
\begin{figure}
	\centering
	\includegraphics[width=0.9\linewidth]{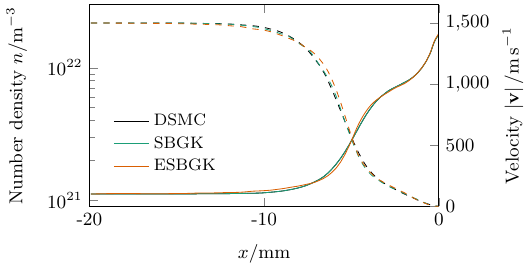}
	\caption{70° blunted cone, Case 1: Velocity in $x$ direction (dashed) and number density (solid) along stagnation streamline using DSMC, ESBGK and proposed SBGK.}
	\label{fig:70degcone-lowdens-1}
\end{figure}

\begin{figure}
	\centering
	\includegraphics[width=0.9\linewidth]{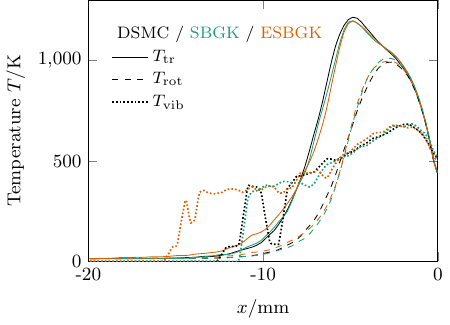}
	\caption{70° blunted cone, Case 1: Translation, rotational and vibrational temperatures along the stagnation streamline using DSMC, ESBGK and proposed SBGK.}
	\label{fig:70degcone-lowdens-2}
\end{figure}

\begin{figure}
	\centering
	\includegraphics[width=0.9\linewidth]{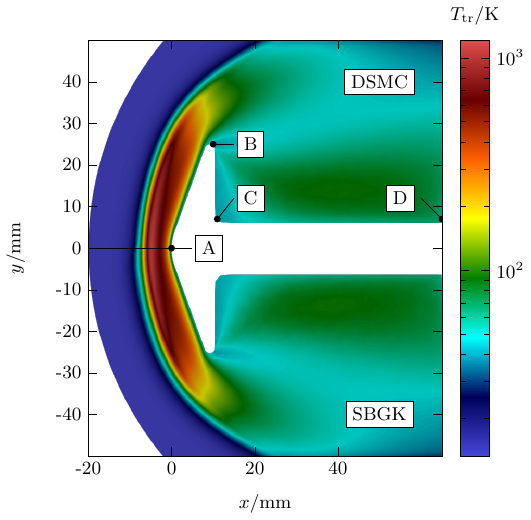}
	\caption{70° blunted cone, Case 1: Comparison of the translational temperature in the flow field using DSMC and SBGK. Characteristic points A-D on the cone surface are indicated.}
	\label{fig:70degcone-lowdens-domain}
\end{figure}

\begin{figure*}
	\centering
	\subfloat{\includegraphics[width=0.45\linewidth]{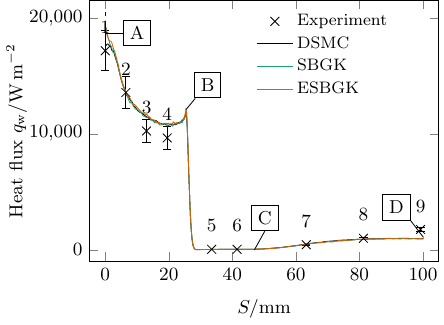}}\,\,
	\subfloat{\includegraphics[width=0.45\linewidth]{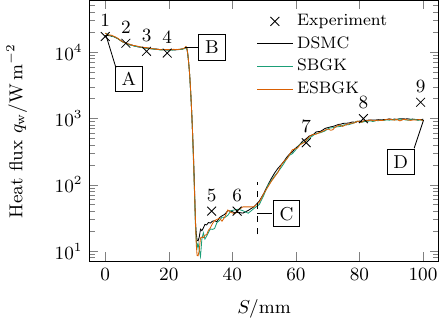}}
	\caption{70° blunted cone, Case 1: Heat flux along the surface with characteristic points A-D indicated (see Figure~\ref{fig:70degcone-lowdens-domain}).}
	\label{fig:70degcone-lowdens-3}
\end{figure*}

The simulation results of the particle number density and the velocity of the N$_2$ flow are shown in Figure~\ref{fig:70degcone-lowdens-1}.
Here, the results of the DSMC simulation are shown as a reference, along with the results of the ESBGK model from~\citet{bgk-poly-mix-hild} based on the same underlying approach.
Overall, the agreement between all methods is quite good. However, the presented SBGK method shows slightly better agreement with the DSMC results than the ESBGK approach.
In Figure~\ref{fig:70degcone-lowdens-2}, the translational, rotational, and vibrational temperatures of the N$_2$ flow are compared between the different methods.
The ESBGK model predicts an earlier onset of the translational temperature rise compared to DSMC, whereas the SBGK method captures the onset of the shock in very good agreement with DSMC.
This behavior is also visible in Figure~\ref{fig:70degcone-lowdens-domain}, where the translational temperature distribution in the flow field is shown for comparison.
This behavior is typical for both the SBGK and ESBGK methods, even in single-species cases without internal degrees of freedom~\cite{piclas-bgk}.
The agreement in the post-shock region is excellent.
Finally, the heat flux along the cone surface is compared in Figure~\ref{fig:70degcone-lowdens-3}.
On both the windward side of the cone and on the rear surface, excellent agreement is observed between all three numerical methods, as well as with experimental measurements.

\subsubsection{Case 2: CO\textsubscript{2}--N\textsubscript{2}}
\begin{figure}
	\centering
	\includegraphics[width=0.9\linewidth]{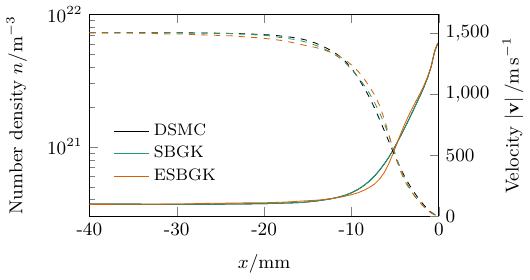}
	\caption{70° blunted cone, Case 2: Mixture mean values of velocity in $x$ direction (dashed) and number density (solid) along stagnation streamline using DSMC, ESBGK and proposed SBGK.}
	\label{fig:70degcone-CO2-N2-1}
\end{figure}

\begin{figure*}
	\centering
	\subfloat[Mean mixture temperatures along the stagnation stream line.]{\includegraphics[width=0.45\linewidth]{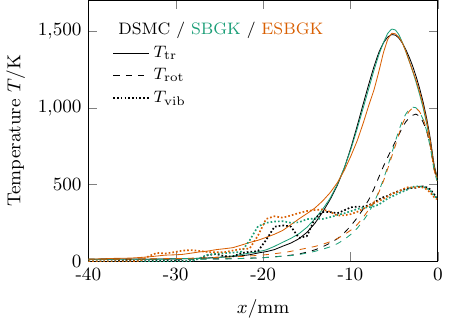}}
	\subfloat[Vibrational temperatures per species at the stagnation stream line.]{\includegraphics[width=0.45\linewidth]{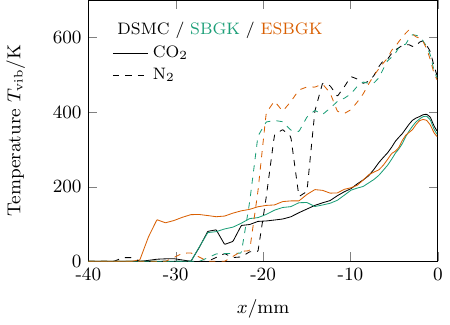}}
	\caption{70° blunted cone, Case 2: Mixture mean values of the translation, rotational and vibrational temperatures and species-specific values of the vibrational temperature along the stagnation streamline using DSMC, ESBGK and proposed SBGK.}
	\label{fig:70degcone-CO2-N2-2}
\end{figure*}

\begin{figure*}
	\centering
	\subfloat{\includegraphics[width=0.45\linewidth]{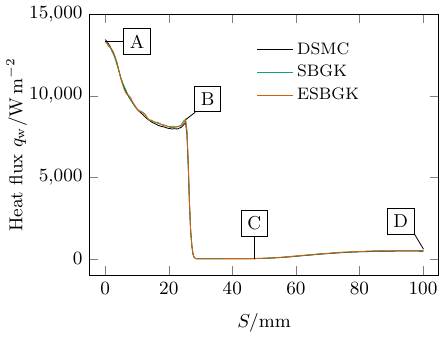}}
	\subfloat{\includegraphics[width=0.45\linewidth]{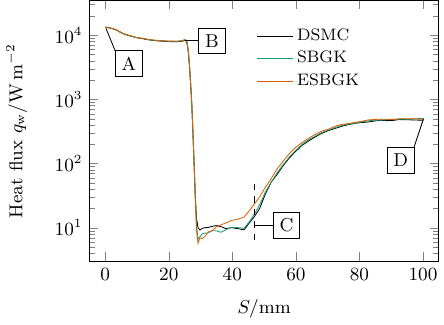}}
	\caption{70° blunted cone, Case 2: Heat flux along the surface with characteristic points A-D indicated (see Figure~\ref{fig:70degcone-lowdens-domain}).}
	\label{fig:70degcone-CO2-N2-3}
\end{figure*}

For Case 2, a 50\,\%\,-\,50\,\% $\mathrm{CO}_2$--$\mathrm{N}_2$ mixture is considered to assess the accuracy of the SBGK model for a polyatomic-diatomic gas mixture with a larger number of internal degrees of freedom, particularly in vibration.
In Figure~\ref{fig:70degcone-CO2-N2-1}, the simulation results for the mean velocity in the $x$-direction and the particle density of the mixture are compared for the DSMC, ESBGK, and SBGK models.
As already observed for the N$_2$ case, the overall agreement is good, with the SBGK results showing slightly better agreement with DSMC than the ESBGK results.
The mean translational, rotational, and vibrational temperatures of the gas mixture are presented in Figure~\ref{fig:70degcone-CO2-N2-2}.
Again, the ESBGK model exhibits a broader shock region due to an earlier onset of the translational temperature rise compared to DSMC, whereas the SBGK method captures both the onset and the shock structure, especially for the translational temperature, in very good agreement with DSMC.
In the post-shock region, both ESBGK and SBGK show very good agreement with the reference solution.
The vibrational temperatures of each species along the stagnation streamline are also shown.
While the agreement between the SBGK and DSMC results is excellent even in the non-equilibrium shock region, the ESBGK model again predicts a slightly earlier increase, particularly for $\mathrm{CO}_2$.
Finally, the comparison of the heat flux along the cone surface in Figure~\ref{fig:70degcone-CO2-N2-3} demonstrates very good agreement of both ESBGK and SBGK with the DSMC reference results.

\subsubsection{Case 3: N\textsubscript{2}--O\textsubscript{2}--N}
\begin{figure}
	\centering
	\includegraphics[width=0.9\linewidth]{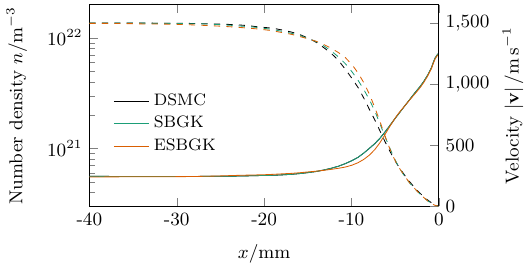}
	\caption{70° blunted cone, Case 3: Mixture mean values of velocity in $x$ direction (dashed) and number density (solid) along stagnation streamline using DSMC, ESBGK and proposed SBGK.}
	\label{fig:70degcone-middledens-1}
\end{figure}

\begin{figure*}
	\centering
	\subfloat[Mean mixture temperatures along the stagnation stream line.]{\includegraphics[width=0.45\linewidth]{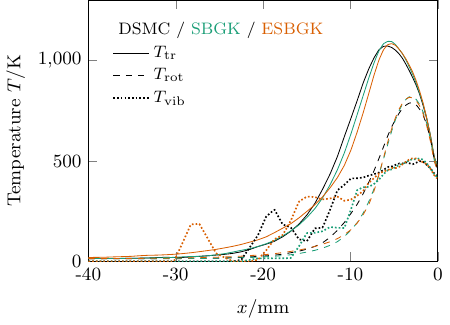}}
	\subfloat[Vibrational temperatures per species at the stagnation stream line.]{\includegraphics[width=0.45\linewidth]{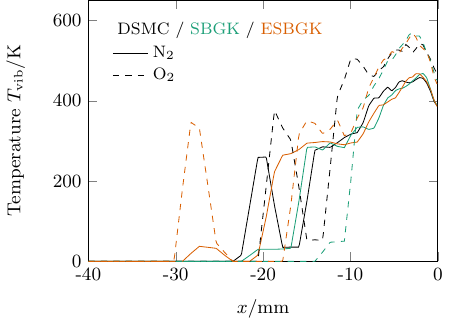}}
	\caption{70° blunted cone, Case 3: Mixture mean values of the translation, rotational and vibrational temperatures and species-specific values of the vibrational temperature along the stagnation streamline using DSMC, ESBGK and proposed SBGK.}
	\label{fig:70degcone-middledens-2}
\end{figure*}

\begin{figure*}
	\centering
	\subfloat{\includegraphics[width=0.45\linewidth]{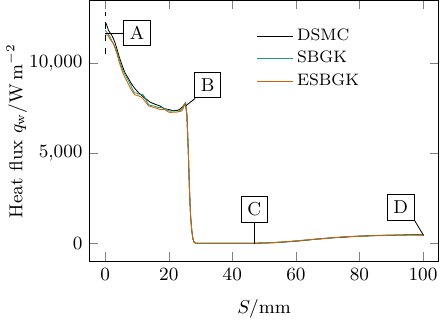}}
	\subfloat{\includegraphics[width=0.45\linewidth]{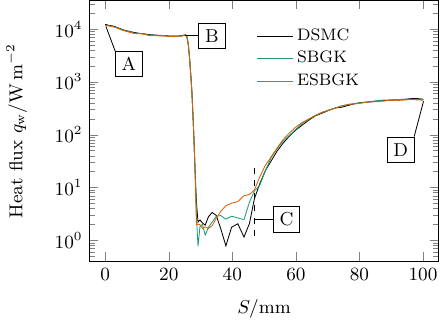}}
	\caption{70° blunted cone, Case 2: Heat flux along the surface with characteristic points A-D indicated (see Figure~\ref{fig:70degcone-lowdens-domain}).}
	\label{fig:70degcone-middledens-3}
\end{figure*}

The simulation results for the mean velocity in the $x$-direction and the particle density of the three-species mixture $\mathrm{N}_2$--$\mathrm{O}_2$--$\mathrm{N}$ are shown in Figure~\ref{fig:70degcone-middledens-1}.
Overall, good agreement is observed between all methods compared to the DSMC reference, again with a slight advantage for the SBGK approach.
In Figure~\ref{fig:70degcone-middledens-2}, the mean translational, rotational, and vibrational temperatures of the gas mixture are compared for the DSMC, ESBGK, and SBGK simulations.
Similar to the previous cases, the ESBGK model predicts an earlier onset of the translational temperature rise compared to DSMC, leading to a broader shock region, whereas the SBGK method captures the shock thickness relatively well.
Excellent agreement is observed in the post-shock region.
Figure~\ref{fig:70degcone-middledens-2} presents the vibrational temperatures of each species.
Although the data exhibit significant noise due to the low inflow temperatures and the quantized treatment of vibration, a good agreement between the methods can still be identified in the statistically relevant regions, i.e., at temperatures above approximately 350~K.
A comparison of the heat flux along the cone surface is shown in Figure~\ref{fig:70degcone-middledens-3}.
Both on the windward side and the rear part of the cone, the ESBGK, SBGK, and DSMC results are in excellent agreement.

\section{Conclusion}
In this work, a Shakhov BGK model is proposed and implemented in the open-source particle code PICLas.
Single molecules and atomic as well as polyatomic mixtures are simulated while accounting for non-equilibrium in the internal degrees of freedom, envisioning a coupling with the DSMC method to solve multi-scale problems.
For verification, a supersonic Couette flow is simulated first with different gas compositions.
Here, good agreement in the temperature results between the SBGK and DSMC methods is achieved with small deviations for an Ar--He mixture, which is assumed to be due to the large mass difference of the atoms, as also seen in previous work.
As a second test case, a flow around a 70$^\circ$ blunted cone is used, exploiting varying inflow parameters and gases.
Overall, very good agreement between the SBGK and DSMC methods is shown for the stagnation steamline and the flow field in front of the body.
The results are also compared with the ESBGK model presented in Ref.~\cite{bgk-poly-mix-hild}, where the most significant difference is visible in the shock region.
While the ESBGK model shows deviations concerning the onset of the temperature increase in the shock region, which is a known behavior for the ESBGK method, the SBGK model captures the shock relatively well compared to DSMC.
The agreement of the simulation results of all three methods in both the post-shock region and on the cone surface, i.e., the heat flux, are excellent.

In future work, the implementation of chemical reactions into the model is envisioned.

\section*{Acknowledgements}
This work is funded by the Deutsche Forschungsgemeinschaft (DFG, German Research Foundation) -- Project-ID 516238647 -- SFB 1667/1 (ATLAS -- Advancing Technologies of Very Low-Altitude Satellites).

\bibliography{sample}

\end{document}